\newif\iflayout
\layoutfalse

\iflayout
   \documentstyle[aps,epsfig,multicol]{revtex}
\else
   \documentstyle[preprint,epsfig,eqsecnum,aps]{revtex}
\draft
\fi

\begin{document}
%\draft

%\twocolumn

%@twocolumnfalse\endcsname
\tightenlines
\title{
Spontaneous magnetization of aluminum nanowires deposited on the 
NaCl(100) surface
}
\author{A. Ayuela, H. R\"abiger, M.J. Puska,  and R. M. Nieminen } 
\address{
Laboratory of Physics, 
Helsinki University of Technology, 02015 Espoo,Finland.
}
\maketitle
\begin{abstract}

We investigate electronic structures of Al quantum wires, both unsupported 
and supported on the (100) NaCl surface, using the density-functional  
theory. We confirm that unsupported nanowires, constrained to be linear, show  
magnetization when elongated beyond the equilibrium length. Allowing
ions to relax, the wires deform to zig-zag structures with
lower magnetization but no dimerization occurs. When an Al wire is deposited
on the NaCl surface, a zig-zag geometry emerges again. The magnetization 
changes moderately from that for the corresponding unsupported wire.
We analyse the findings using electron band structures and simple
model wires. 
%Our results aim to encourage experiments searching for magnetism of 
%simple metal nanowires deposited on inert surfaces.
\end{abstract}
\noindent {\it PACS \# \ \ }

%\vskip2pc]
\iflayout
   \begin{multicols}{2}[]
\narrowtext
\fi

\section{Introduction}

Electronic properties of nanowires adsorbed on solid surfaces depend 
strongly on the geometric structure of the wire. The surface enables
the realization of a specific geometry but the interaction
between the substrate and  the adsorbed wire 
does not necessarily destroy the key properties the wire would have 
in the same geometry but as isolated. This idea has been the basis of 
many recent experimental and theoretical studies of nanowires 
on semiconducting \cite{watanabe,segovia,yeom,lopinski,losio} 
and on metallic \cite{pampuch,bazhanov} substrates. The
studies of nanowires on surfaces have been motivated from the 
new technology point of view. Nanowires can be seen as ultimate
leads when miniaturizing devices in electronics. Moreover, if they
exhibit magnetization they could be used as compact magnetic storage
devices. In general, the exciting quantum phenomena taking place
as the size and the dimensionality of the system are reduced
are the driving force in the research of  nanostructures.

The manufacturing of nanowires on solid surfaces can be based
on the manipulation of the surface and the adsorbate atoms with
the scanning tunneling microscope (STM) \cite{eigler,shen}. 
A more efficient method for fabrication may be the
deposition of adsorbate atoms in the steps on metal surfaces
\cite{gambardella,dallmeyer} or the growing of stripe-like 
adsorbate structures either on metal \cite{pampuch,koh} or on semiconductor
\cite{segovia,lopinski,losio} surfaces. The geometry of these structures
can be monitored using STM \cite{nielsen,lopinski,gambardella}, 
He scattering \cite{gambardella} or by field ion microscope \cite{koh}. 
An efficient way to characterize the electronic structures of wires, 
{\em e.g.} their one-dimensionality, is the angle-resolved photoemission
\cite{segovia,losio,pampuch}.  

In this work we report state-of-the-art electronic structure calculations 
based on the density-functional theory (DFT) for
simple metal nanowires on an insulating substrate, {\em i.e.} for 
Al wires on the NaCl(100) surface. The present work is largely inspired
by the theoretical investigations of Ga adsorbate wires on Si(100)
surface performed by Watanabe {\em et al.} \cite{watanabe2}. Using 
first-principles electronic structure methods they calculated
the stable structures of Ga - dangling-bond wires on hydrogen-terminated
Si(100). The resulting bandstructures show that the wires are
metallic and that in some of the structures the adsorbate-induced
band near the Fermi level is very flat. On this basis Watanabe {\em et al.} 
\cite{watanabe2} discussed, using a simple tight-binding model, the
possibility of the ferromagnetism of these simple metal wires. 
Recently, Okada and Oshiyama \cite{okada} returned to these Ga wires
on Si(100) performing first-principles DFT calculations. They found one 
atomic configuration for the Ga wire  exhibiting ferromagnetic ordering.   
We have chosen to work with Al wires because the jellium model
calculations \cite{zabala-m} have predicted, and the subsequent 
first-principles calculations \cite{aay} confirmed, that isolated 
linear Al atom chains can be magnetic for certain interatomic 
distances. NaCl has been selected as the substrate due to its 
insulating character, which is expected to be reflected as minor
substrate effects in the electron structure of the Al wire, and due to
the fact that the atomic distances on the NaCl(100) surface match with
the interatomic distance of a magnetic isolated Al-chain. This surface
is also expected to be stable, as a surface reconstruction occurs only in one 
of many different cleavages \cite{glevov}. An Al wire on the unreconstructed
NaCl surface is a much simpler system than Ga wires on
the hydrogenated reconstructed Si(100). Therefore the present calculations
are expected to show clearly the essential physical phenomena for
simple metal wires on insulating surfaces. Because the
properties of a nanowire on a surface are largely derived from those of
unsupported wires, we begin this work by calculations for isolated wires 
in different geometries and study their properties as functions of 
structural parameters. 

The organization of the present article is as follows: In Sec. \ref{sec:theory}
we discuss the computational details. We use the Vienna {\it  ab initio}
simulation  package  (VASP)  \cite{vasp} which is a 
pseudo-potential-plane-wave (PPPW) code. Moreover, we have made benchmark 
calculations using the the full-potential linearized augmented plane wave 
(FLAPW) method with the WIEN97 package \cite{WIEN97}. FLAPW is considered more 
accurate but it is also much more cpu-time-consuming than the 
PPPW method. In Section III  we present our results,
first for the unsupported linear (III A) and dimerized or zig-zag wires 
(III B) and thereafter for wires on the NaCl(100) surface (III C). 
Section IV summarizes our results, and includes an outlook.

%Our aim
%is to encourage experiments searching for magnetism of simple metal nanowires  
%deposited on inert surfaces.

\section{Computational details}
\label{sec:theory}

We employ in our DFT electronic structure calculations
the generalized gradient approximation (GGA) for the 
exchange-correlation effects \cite{perdew-gga}. 
%The local density approximation (LDA) is employed for the 
%exchange-correlation effects \cite{perdew-lsd}. 
When describing interactions between adsorbate atoms
and a substrate it has been found important go beyond the
local density approximation (LDA). We do both  
spin-polarized and spin-compensated  computations in  order to 
find the  solution of the lowest energy. 

In the PPPW calculations (VASP package) 
we have used ultrasoft  Vanderbilt pseudopotentials \cite{us-pp}.
In the case of unsupported Al-wires the cutoff energy of the
plane-wave expansion has been 180 eV. This is sufficient to describe 
bulk Al. In the case of the NaCl surface we need a cutoff energy of 
250 eV. 

We use the supercell approximation in our calculations. This means that 
instead of isolated atomic chains a set of parallel chains are used. 
In the case of linear chains
the interchain distance of 15 \AA \ is sufficient to eliminate the
interaction between the wire and its periodic replicas. The zig-zag
chains are treated in a supercell with the lattice constant of 20 \AA \  
perpendicular to the chain axis. In the NaCl(100) surface calculations 
we have used the slab geometry with one or two layers of Na and Cl atoms. 
The use of one layer is found sufficient to describe the energetics
of straight Al wires on the surface. The NaCl slabs are separated
by a vacuum region of 20 {\AA}. On  the surface, the Al wires are 
separated by 3 lattice constants of NaCl. These separations were
found to depress sufficiently the interactions between the slabs or
the Al wires. Dipole corrections for the electrostatics of the 
supercell geometry are also included \cite{makov}.

The desired periodicity along the wire and the artificial periodicity
due to the supercell approximation require a Brillouin zone 
integration. In order to describe correctly the magnetic properties
the {\bf k}-point sampling in the wire directions has to be dense
enough. We have used  26 {\bf k}-points for this purpose. 
In the case of unsupported wires there are {\bf k}-points
only on the $k_z$ axis in the wire direction. When calculating Al
wires on the NaCl(100) surface the {\bf k}-points form a 3x26 mesh  
on the $(k_x, k_z)$ plane  parallel to the surface and there are 
actually 26 {\bf k}-points in the irreducible wedge of the Brillouin 
zone. The Fermi-smearing of 0.06 eV is used in the integrations.

In order to test the accuracy of the PPPW calculations, especially 
with respect to the predicted magnetic properties, we have performed 
calculations also with the FLAPW method  (WIEN97 code) for the unsupported 
linear Al atom chain. The FLAPW method can be considered as one of the 
most accurate band structure methods. In the calculations,
the  muffin-tin sphere radii for Al is  2.10 a.u. Inside the
muffin-tin spheres the maximum angular momentum $l$ in the radial expansion is
$l_{max}=10$, and the largest $l$  value for the non-spherical part of
the Hamiltonian matrix is  $l_{max,ns}=4$.  The cutoff parameters are
$RK_{max}=9$ for  the plane  waves, the number  of plane  waves ranges
up to around 4000.  The Brillouin zone integrations are done using  10 special
points \cite{mp}, and a Fermi broadening of 0.001 Ry is used.

\section{Results}
\label{sec:results}
\subsection{Unsupported linear chain of Al atoms}

First we want to test the accuracy of the pseudo-potential approach
by comparing the PPPW and FLAPW results in the simple case of
an unsupported linear chain of Al atoms. The total energies 
$E_{tot}$ per atom are shown in the uppermost panel of Fig. \ref{AlEtot} 
as a function of the bond length. The energy minima correspond to 
spin-compensated solutions. According to the FLAPW method the energy  
minimum occurs  at $d_{nn}=$  2.43  \AA \ which is in a fair agreement 
with the PPPW result of 2.41  \AA.  The equilibrium force constants 
$k = {d^2 E_{tot} \over  da^2}$ are 104 N/m (FLAPW) and 151 N/m (PPPW).
Our GGA-PPPW values are quite close to the LDA results obtained by Mehrez 
and  Ciraci \cite{mehrez} using a PPPW method. According to 
Fig. \ref{AlEtot} the difference between the FLAPW and PPPW results 
increases towards to the larger interatomic distances. This reflects 
some difficulties in the transferability of the pseudopotential for 
larger distances and low dimensionality, because the increase of the 
energy cutoff does not change the PPPW results substantially.

\begin{figure}[hbt]
\begin{center}
\epsfig{file=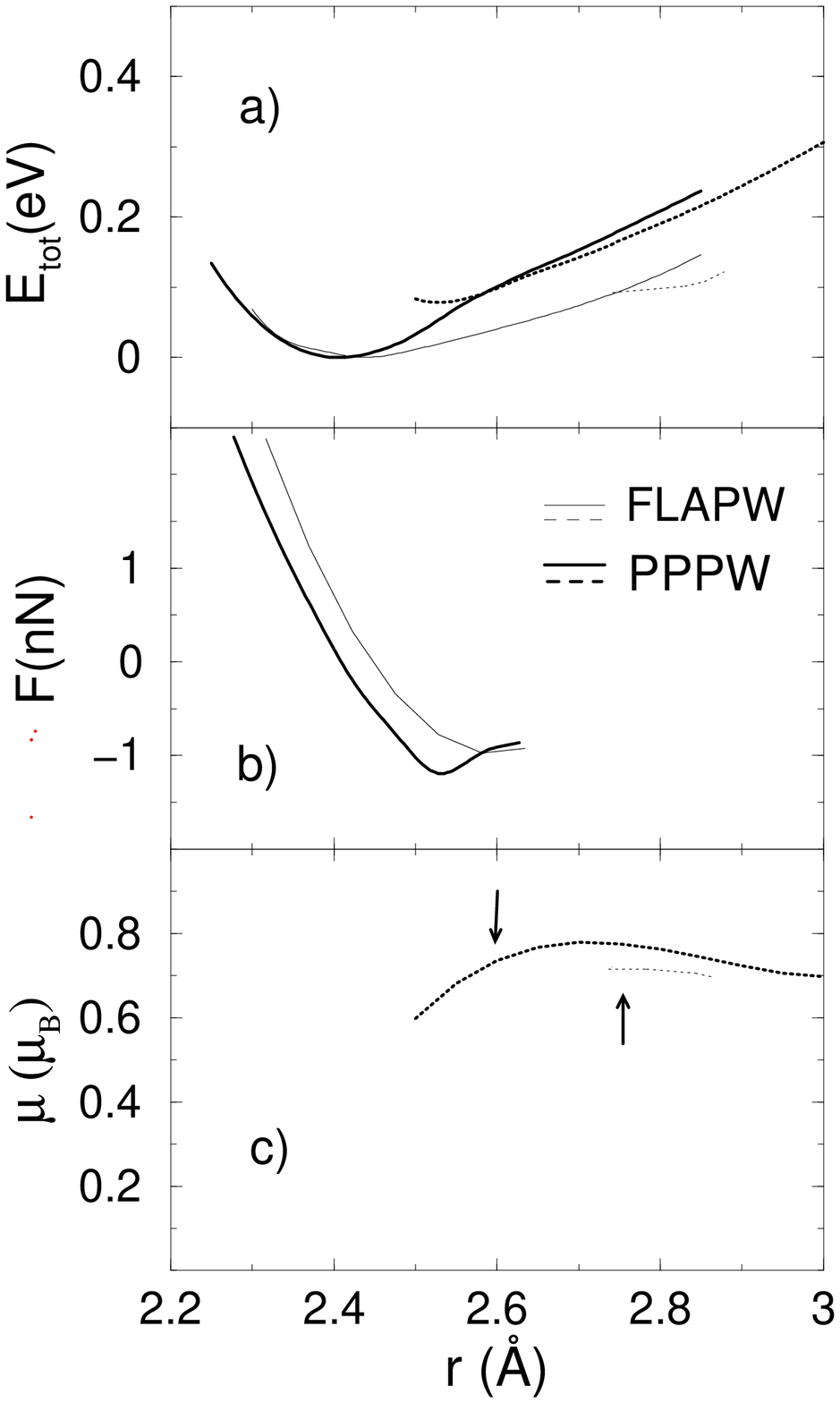,width=0.8\linewidth}
\caption{Total energy (a), elongation force (b), and magnetic moment (c)
per atom for an infinite linear chain of Al atoms as a function of the
interatomic distance. The FLAPW  and PPPW results are denoted by thin and
thick lines, respectively. In panel (a), the minimum energy 
defines the energy zero and the results of spin-compensated 
and spin-polarized calculations are given by solid and dashed
lines, respectively. The arrows in panel c) indicate points where 
the spin-polarized states become in elongation more stable than
the spin-compensated ones.}
\label{AlEtot} \end{center}
\end{figure}

The stabilized-jellium model calculations have predicted that nanowires of
simple metals may be magnetic \cite{zabala-m}.
In Fig. \ref{AlEtot} the results obtained using the spin-compensated
formalism are connected with solid lines whereas solutions with
magnetic moments are connected with dashed lines.
At distances  larger than  2.7 \AA \ (FLAPW) or 2.6  \AA \ (PPPW) the
total energy of a magnetic  solution is indeed lower than that of the
spin-compensated  one. The magnetic solution is due to the  spin
polarization of the second subband accommodating atomic 3$p$ electrons
(see  Figure \ref{Albands}). In the middle panel of Fig. \ref{AlEtot}
the resulting magnetic moment per atom is given as a function of
the elongation. The magnetic moments calculated by both methods saturate
to a value around 0.7 $\mu_B$ per Al atom. It is interesting 
to compare this value with the stabilized-jellium model prediction which
gives, corresponding to the interatomic distance of 2.86 \AA \ 
(bond length in bulk Al), a moment of 0.65 $\mu_B$ per Al atom
\cite{zabala-e}.

The elongation force is defined as the force opposing the lengthening
of the wire due to an external applied force. When negative, the 
elongation force would like to shorten the wire, whereas a positive 
elongation force without a counterbalancing force would lead to a spontaneous 
elongation. In the case of a linear atomic chain the elongation force 
is calculated as
\begin{equation}
F = - \frac{dE_{tot}}{dr},
\label{force}
\end{equation}
where $r$ is the interatomic distance. Naturally, the force vanishes
at the minimum energy and is negative for larger distances.
For very large distances the force should approach zero. 
The crucial point is that when the wire is stretched, the force 
reaches a minimum which determines the maximum force sustained before 
breaking. In both calculations, the breaking force is about 1.5 nN.
This is of the same magnitude as the typical force before breaking
seen in the atomic force microscope experiments \cite{rubio} on
gold nanowires. According to Fig. \ref{AlEtot} the breaking of the 
linear Al chain occurs as a function of the elongation before the 
magnetization sets on. If the chain would be stable, the onset of the 
magnetization should result in a discontinuous decrease of the 
magnitude of the elongation force \cite{zabala-m}.

It is instructive to compare the one-dimensional bandstructure of
the linear Al chain with the properly folded subband structure
of an infinitely long stabilized jellium cylinder. This is done
in Fig. \ref{Albands}. The Al chain is calculated using FLAPW
for the interatomic distance of 2.86~\AA. \
The splitting of the bands due to the magnetic ground state
occurs at this distance. The PPPW method gives quantitatively similar 
results. In the stabilized-jellium calculation\cite{zabala-e} 
the density $n$ of the positive jellium charge corresponds to the
electron density parameter $r_s = (3/4\pi n)^{1/3}$ = 2.07 a.u. The 
radius, 1.35 \AA, \ of the positive background charge is chosen to give the 
linear charge density of 3 $e$/ 2.86 \AA. \ The electronic structure 
has been solved self-consistently within the LDA. The above value 
of the jellium radius falls in a narrow window, in which the
the second, doubly-degenerate subband, corresponding to the quantum 
numbers $m$~=~1 and $n$ = 1, is totally spin polarized resulting in the 
magnetic solution. In the atomic chain calculations the corresponding, 
nearly totally polarized band arises from the atomic $p_{xy}$  orbitals
(the $z$ axis is  parallel to the chain). In a quasi-one-dimensional system 
the density of states (DOS) diverges at the bottom of a subband. When
the Fermi level is just above the bottom of a subband, the Fermi-level DOS
will be large and the system will fulfill the Stoner criterion for
magnetism \cite{zabala-m}. The quantum numbers of the lowest jellium 
subband are $m$~=~0 and $n$ = 1. The  corresponding atomic chain  states 
have the character of the atomic 3$s$ orbital on the lowest branch and 
after the band is folded at $k_z = \pi/a$ the character is that of the 
atomic 3$p_z$ orbitals. The contribution of the lowest subband to the
magnetic moment of the wire is small in comparison with the second
(nearly) totally polarized subband. The  agreement between the jellium
and atomic chain models is  surprisingly good. This gives credence
for  the stabilized jellium model for Al even in the present very 
confined geometry.

\begin{figure}[hbt]
 \begin{center}
\epsfig{file=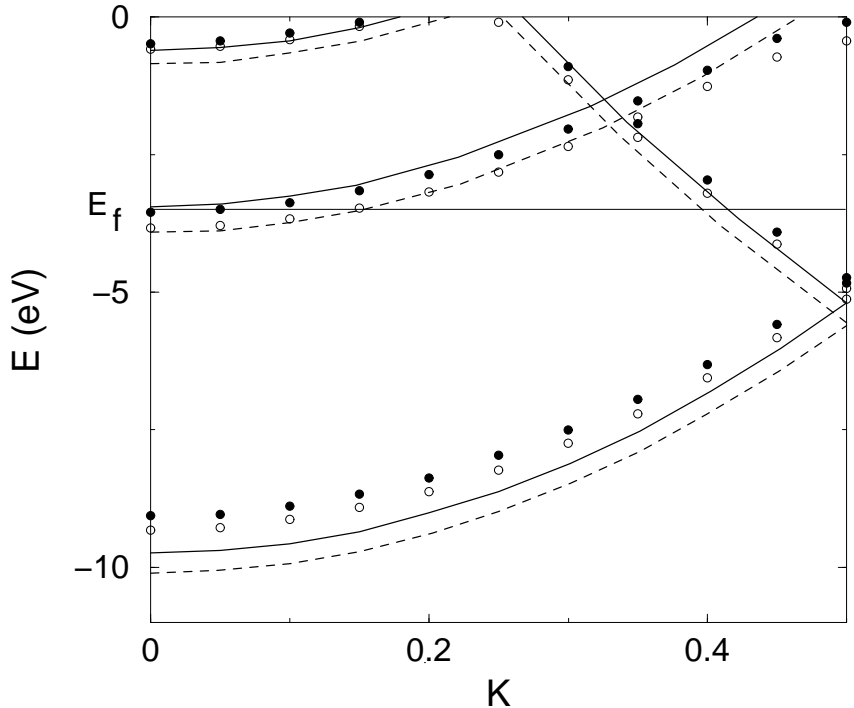,height=0.6\linewidth}
\caption{Band  structure of an unsupported linear chain of Al atoms.   
The interatomic  distance is the Al bulk bond distance of 
$d_{nn}=$ 2.86 \AA. \ The results of the  FLAPW  calculation
are denoted by open and filled circles for the majority and 
minority-spin bands, respectively. The properly folded subband
structure of an infinitely long stabilized-jellium wire with
$r_s$ = 2.07 a.u. and radius of 1.35 \AA \ is shown as dashed
and solid lines corresponding to the majority and minority spins,
respectively. The Fermi level is given by the horizontal thin line.
}  
\label{Albands} \end{center}
\end{figure}

In conclusion, in the case of  the unsupported linear chain of Al atoms 
the PPPW calculations are able to reproduce the essential cohesive properties 
obtained in the all-electron FLAPW calculation. Also the magnetic properties, 
which are of a especial interest in the preset work, are similar in both 
calculations. Therefore in the following calculations
for the unsupported dimerized and zig-zag Al wires as well as for the
Al wires on the NaCl (100) surface we rely on the less
computer intensive PPPW method.

\subsubsection{Unsupported dimerized and zig-zag Al wires}

The next step in our study is to allow the atoms of an unsupported
linear Al chain to dimerize and also to form zig-zag wires.
These geometries are studied because a linear chain of atoms
is susceptible to Peierls-like distortions, opening up a small 
band gap and lowering thereby the total energy. The notations of 
the geometries studied are presented in Fig. \ref{w-geom}. In the
dimerized geometry the distance $2r$ between every  second atom remains 
constant, but the atom in the middle of a cell moves in the direction of
$r_{short}$ along the wire. In zig-zag wires the nearest neighbor distance 
$r$ is kept constant while the zig-zag angle $\theta$ is varied. When
studying the distortions of the spin-polarized chain the distance $r$ 
is 2.8 {\AA} whereas for the spin-compensated chain the linear equilibrium
value of 2.41 {\AA} is used.

\begin{figure}[hbt] 
\begin{center}
\epsfig{file=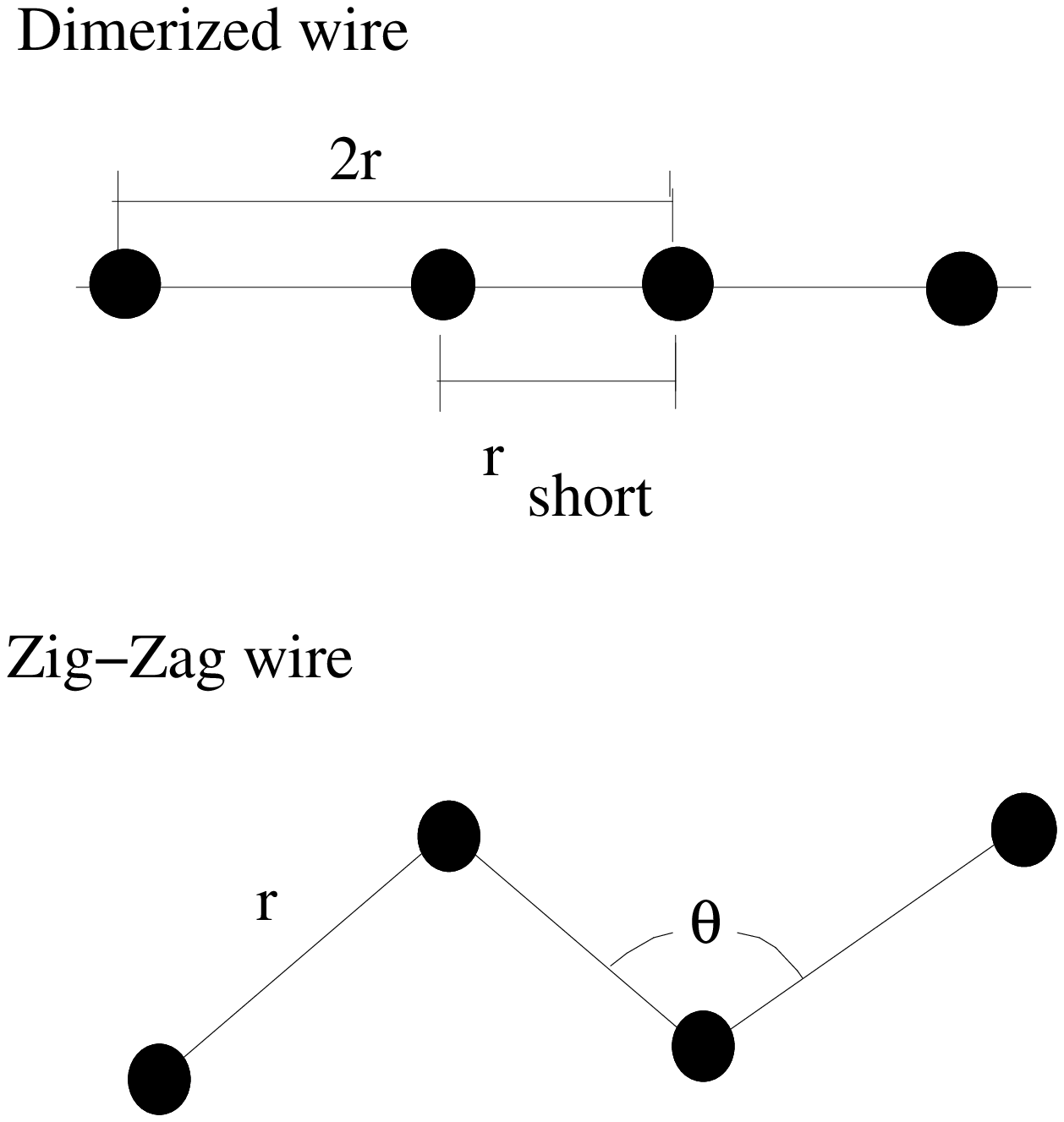,width=0.7\linewidth}
%\psfragscanon 
% \psfrag{theta}{$\theta$} 
   \caption{{ Geometries of the unsupported dimerized and zig-zag wires}}
\label{w-geom} \end{center}
\end{figure}

According to Table \ref{dimers}, the energy of the Al wire increases with
dimerisation, {\em i.e.}, the bond length alternation for the Al chain
vanishes or is at least below our error estimate of around 2~\% of the
interatomic distance. We have actually confirmed this energy increase
also with the  FLAPW  calculations.  The energy  
increases more rapidly close to the equilibrium distance corresponding
to the non-magnetic wire than  close to the larger distance corresponding
to the magnetic wire. It is logical to expect that changes in the geometry
of an wire already stretched have a smaller  
effect  on the  energy  than the changes close  to the equilibrium.
One can also note in the case of spin-polarized wires that
the magnetic  moment grows slightly along dimerization.

\begin{table}[hbt]
\caption{{Energies and magnetic moments of dimerized Al wires. 
The minimum energy for the linear chain of Al atoms defines the
energy zero. The 
spin-compensated and spin-polarized wires are distorted relative to
the linear chain of Al atoms with the bond lengths of 2.39~\AA \ and
2.8~\AA, respectively.}}
\label{dimers}
\begin{center}
\begin{tabular}{|l|c||l|c|c|}
\multicolumn{2}{c}{Spin-compensated} & \multicolumn{3}{c}{Spin-polarized} \\
\hline
$r_{short}$ (\AA)  & $E (eV)$  & $r_{short}$ (\AA)  & $E$ (eV)  & $\mu
(\mu_B)$ \\
\hline
2.41 &   0    & 2.80 &  0.192 & 0.76 \\
2.35 &  0.007 & 2.75 &  0.189 & 0.76 \\  
2.30 &  0.036 &  2.7 &  0.195 & 0.77 \\ 
2.25 &  0.085 & 2.65 &  0.206 & 0.78 \\ 
2.2 &   0.160 &  2.6 &  0.221 & 0.80 \\
\hline
\end{tabular}
\end{center}
\end{table}

Next we consider zig-zag wires depicted in Fig. \ref{w-geom}. 
Table \ref{zig-zag} gives the total energies and magnetic moments 
obtained.  The zig-zag formation is energetically favorable
at the distance $r$ = 2.41~\AA \  corresponding to spin-compensated 
solutions as well as at $r$ = 2.80~\AA \  giving magnetic solutions. 
The values of the equilibrium zig-zag angles $\theta$ are around 
150$^o$ and 130$^o$ for the spin-compensated and the magnetic 
systems, respectively. The deformation is limited by the 
second-nearest-neighbor atom interactions in the atomic chain. 
In the magnetic chain, the longer bond length $r$ allows for a angle 
deformation larger than in the spin-compensated chain. 

\begin{table}[h]
%\begin{table}[hbt]
\caption{{ Energies and magnetic moment of zig-zag Al wires. 
The minimum energy for the linear chain of Al atoms defines the
energy zero. The 
spin-compensated and spin-polarized wires are distorted relative to
the linear chain of Al atoms with the bond length of 2.39 \AA \ and
2.80 \AA, \ respectively
}}
\label{zig-zag}
\begin{center}
\begin{tabular}{|l|c||l|c|c|}
\multicolumn{2}{c}{Spin-compensated} & \multicolumn{3}{c}{Spin-polarized} \\
\hline
{$\theta$} & {$E (eV)$} & {$\theta$} & {$E (eV)$} & {$\mu (\mu_B)$} \\
\hline
180 &  0     & 180 & 0.192 & 0.76  \\ 
170 & -0.008 & 170 & 0.178 & 0.69 \\  
160 & -0.023 & 160 & 0.151 & 0.54  \\ 
150 & -0.032 & 150 & 0.101 & 0.25 \\ 
140 & -0.019 & 140 & 0.058 & 0.28 \\ 
130 & +0.021 & 130 & 0.046 & 0.25 \\ 
    &        & 120 & 0.067 & 0.31 \\ 
    &        & 110 & 0.076 & 0.24 \\
\hline
\end{tabular}
\end{center}
\end{table}

According to Table \ref{zig-zag} the energies of the of the chains
decrease when the zig-zag angle decrease from 180$^o$. Especially
for the magnetized chain the decrease is strong and it is accompanied 
by a decrease in the magnetic moment. The minimum values are
obtained at  the angle  of 120$^o$. To understand this behavior an 
analysis of the band structure required. We do this analysis here, 
in the context of the unsupported Al chains, because this analysis will 
be the basis to understand the magnetism of the more complicated systems 
of Al wires on the NaCl(100) surface. The  band structure  of the magnetic 
zig-zag wire at  the equilibrium zig-zag angle of 130$^o$ is plotted in 
Figure \ref{bs130} a). It is dramatically different from that for
the linear chain in Fig. \ref{Albands}. As the zig-zag angle $\theta$ 
decreases, the width of  the {\it s}-band decreases and the  band gap at 
around -5 eV  increases. The states of the 3$p$-bands that cause 
the magnetization undergo also significant changes. There is no longer  
a doubly-degenerate $p_x-p_y$ band.  
Two $sp^2$-like bands are formed, the lower of which is occupied both 
by spin-up and spin-up  electrons whereas the upper $sp^2$-like band
is well above the Fermi level. A totally polarized $p_z$-band causes the 
magnetization. As one  third of  the Al 3$p$-electrons occupy the 
$p_z$ band, a magnetic moment  of roughly 0.3 $\mu_B$ can be expected
from this interpretation. This value is indeed in agreement with 
the actual computed result in Table \ref{zig-zag}. At the  zig-zag angle 
of  120$^o$ (see Fig.  \ref{bs130} b) ) the lower $sp^2$-like band 
becomes polarized at low {\bf k}-values. This explains the  increase of the 
magnetic  moment at small zig-zag angles. 
%(In Fig. \ref{bs130}  b) for $\theta$ = 120$^o$ the Fermi level splits
%the spin-down and spin-up bands). 

\newpage

\begin{figure}[hbt]
\begin{center}
\epsfig{file=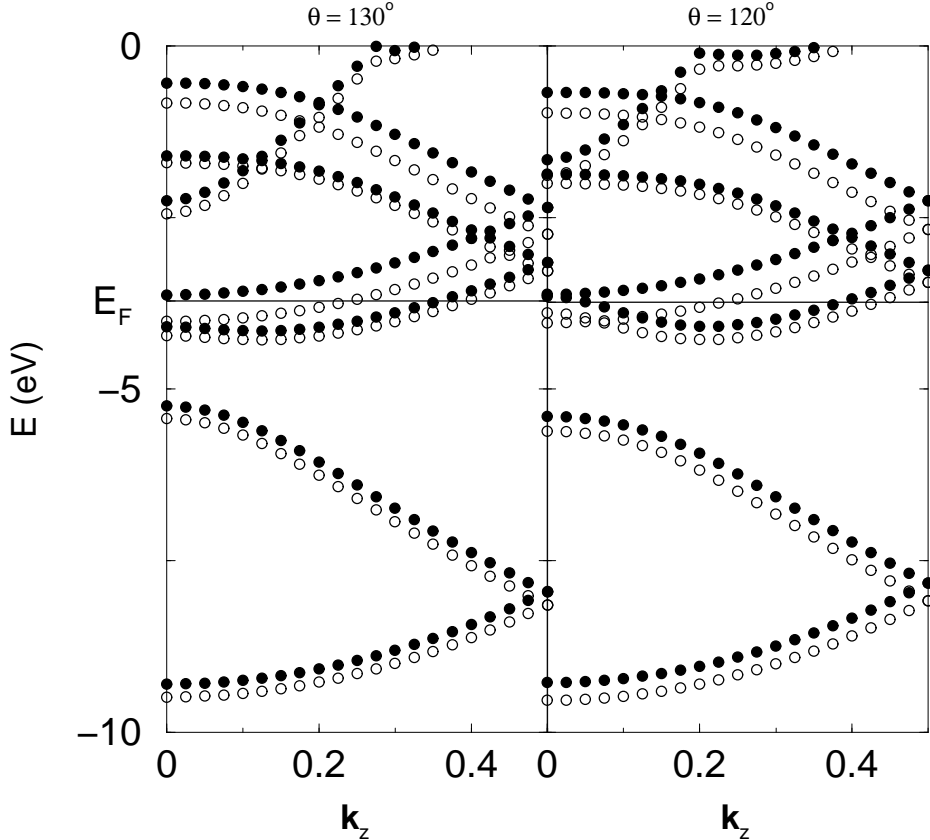,width=0.7\linewidth}
\vspace{0.5cm}  
\caption{{ Band structure of an unsupported magnetic
Al nanowire with the zig-zag angle of a) $\theta$ = 130$^o$ and 
b) $\theta$ = 120$^o$. The nearest  neighbor distance is 2.8 \AA. 
Spin-up  and spin-down bands are denoted by open and filled circles, 
respectively. The Fermi level is given by the horizontal thin line.
}}  
\label{bs130} 
\end{center}
\end{figure}

\subsection{Al wire on the NaCl(100) surface}

In this subsection we present our results for an Al wire in
equilibrium on the NaCl(100) surface and at the end of
of the subsection we analyze these results using simple model
calculations. The NaCl(100) slabs used in the calculations
correspond to the atomic layers cut from the perfect bulk
crystals with the theoretical lattice constant of 5.68 \AA .
The omission of the relaxations of Na and Cl ions is justified
because we are interested mainly in the effect 
the surface has on the wire and not vice versa.
Before dealing with Al wires on the surface we have made
calculations for single isolated Al atoms adsorbed at 
different sites on NaCl(100). The sites on the top of 
Na and Cl ions, on the middle of the bridge between the two ions
as well as on the hollow site in the center of four ions
have been considered. These calculations show that the
properties of the adsorbed Al atoms (adsorption energy and the
distance from the surface) do not change remarkably when
three NaCl(100) layers are used instead of two or the surface
supercell is increased from 3x3 to 4x4. The following
results correspond to two NaCl(100) layers and the 4x4 cell or
in the case of the Al wire, the interwire distance is three
NaCl lattice constants.

The results for the isolated Al atom on NaCl(100) serve also as 
guidelines for later calculations dealing with the Al chains on 
the NaCl(100) surface. The optimized adsorption heights from the 
surface-layer atoms and the corresponding total energies are given 
in Table \ref{tab:surfa}. The isolated Al atom prefers to be on the 
top of a Na atom. The total energies are therefore given
relative to this system. Thereafter the next favorable site is
the hollow one followed by the bridge site. The adsorption of Al on
the top of the Cl atom is clearly the most unfavorable case.

\begin{table}[hbt]
\caption{{ Adsorption of a single Al atom on the NaCl(100) surface.
The height $d$ of the adsorbed Al atom from the surface atom layer 
and the total energies $E$ relative to that for an Al atom on the top 
of a Na atom are given.}}
\label{tab:surfa}
\begin{center}
\begin{tabular}{|c|l|c|}
position     & d(\AA)        & {$E (eV)$}  \\
\hline
 top Na &  2.73 & 0.00 \\
 middle  face & 2.43 & 0.06 \\  
middle bond & 2.83 & 0.16 \\ 
top Cl & 3.83 & 0.32 \\
\hline
\end{tabular}
\end{center}
\end{table}

The results for the isolated Al on NaCl(100) in mind we start 
the simulations for the Al wire commensurate on the NaCl(100) surface.
In these calculations there are two rigid layers of Na and Cl ions
and the supercell contains two Al atoms. The equilibrium positions of 
Al atoms obtained reflect the Al-atom - surface
interaction as well as the Al-Al interactions within the wire.
It turns out that wires in which the Al atoms sit on the top of 
nearest Na atoms along the [110] direction as straight or zig-zag 
chains have not the lowest energies. Also, when we start (inspired 
by the results for unsupported wires) the structure optimization 
from an initial configuration, in which the Al atoms sit close to 
the adjacent hollow sites forming a slightly zig-zaged chain 
a strong structural relaxation takes place. The resulting configuration
shown in Fig. \ref{s-geom} a) is a  zig-zag  wire in which the 
Al atoms reside in the bridges between the Na and Cl atoms.  
The Al atoms are closer to the Cl atoms than the Na atoms so that the
zig-zag angle $\theta=143^o$ and the interatomic distance in the chain is
$r$ = 2.99 \AA . The distance to the surface is $d$=  2.90 \AA.
It is exciting that the Al wire in this equilibrium configuration has
a magnetic moment of 0.24 $\mu_b$ per atom! This is just of the
order of magnitude expected from Table \ref{zig-zag} for unsupported
zig-zag wires. Thus the interaction with the NaCl surface does not
destroy the magnetic moment of the quasi one-dimensional Al wire.

The above results underline the importance of relaxing the positions
of the wire atoms on the surface. This kind relaxation should be
performed also when studying the early low-coverage stages of the
growth of metal layers on insulating surfaces \cite{fuks}. The
approximation that lateral interactions between adjacent adsorbate
atoms are unimportant is maybe not valid one for all growth patterns.

In order to understand the properties of the above equilibrium wire 
we have made further calculations using simplified models for linear
Al wires on the NaCl(100) surface. In these calculations only one 
rigid layer of NaCl has been used but the distance between the adjacent
wires due to the supercell structure is the same as before. 
The Al atoms are allowed to relax only perpendicular to the surface.
The geometries considered are presented and labeled in Fig. \ref{s-geom}. 
An Al atom has in these geometries one (NN1), two (NN2) or four 
(NN4) surface ions as nearest neighbors corresponding to atop, bridge
and hollow site wires. The supercells in these 
simulations contain 12 non-equivalent surface ions and two wire atoms.

\begin{figure}[hbt]
 \begin{center}
  \epsfig{file=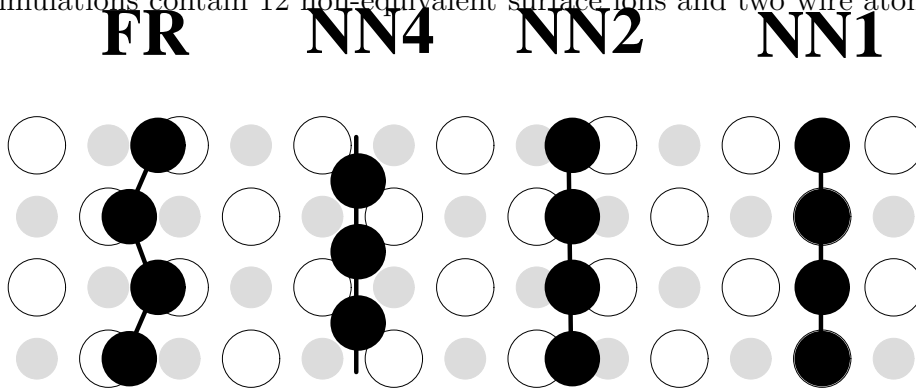,width=0.7\linewidth}
\vskip 15pt
  \caption{{Geometries of commensurate Al wires on the
   NaCl(100) surface. The top view of a fully relaxed wire (FR) as
   well as those of model wires with one (NN1), two (NN2), and
   four (NN4) nearest neighbor surface atoms are given. The small grey and  
  large white circles denote the Na and Cl atoms, respectively.}}
\label{s-geom} 
\end{center}
\end{figure}

The energies of the model wires calculated per one Al atom are shown in 
the lower panel of Fig. \ref{w_surf_curves} as a function of the distance
from the surface layer. The NN2 wire has the lowest equilibrium energy
and this is used to define the energy zero. The equilibrium distance
is nearly the same as for the fully-relaxed zig-zag wire discussed 
above. The NN4 wire has a clearly higher energy. This explains the 
behaviour of our above-discussed simulation starting from Al atoms near 
the hollow sites. The second lowest  energy minimum  is obtained  in the  NN1  
geometry, in which the Al adjacent Al atoms sit on the top of both
Na and Cl surface atoms. Clearly, this geometry in which both Al atoms
of the supercell have the same distance from different types of 
surface atoms is physically not meaningful. Anyway, the fact that
the minimum total energy of the NN1 wire is lower than that of the
NN4 wire further supports the unstable character of the NN4 wire.
In conclusion, the comparison of the results of these model 
wires with those for single isolated Al atom on the NaCl(100)
surface reflect the balance between the Al-surface interactions 
and the intrawire interactions. 

\begin{figure}[hbt]
 \begin{center}
  \epsfig{file=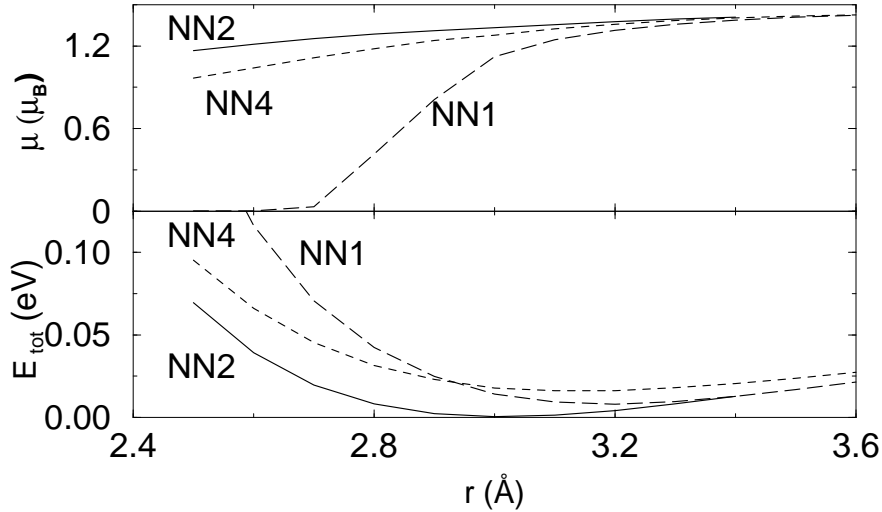,width=0.7\linewidth}
  \vspace{0.5cm} \caption{{Commensurate Al wires on NaCl(100).
The total energy (lower panel) and the magnetic moment (upper panel) per Al 
atom are shown as functions of the wire distance from the surface. The 
long-dashed, solid, and short-dashed curves correspond to the
geometries NN1, NN2, and NN4, respectively (See Fig. 
{\protect \ref{s-geom} }). The minimum energy of NN2 determines
the energy zero.}}    
\label{w_surf_curves}
  \end{center}
\end{figure}

The upper panel of Fig. \ref{w_surf_curves}
shows that all the model wires are magnetic over a wide region around
the equilibrium distance. It is interesting to note that the magnetism
of the NN1 wire vanishes when the distance to the surface decreases.
In order to understand the magnetic properties of these wires as well
as those of the fully-relaxed wire on NaCl(100) we need to analyze 
their band structures. The equilibrium band structure of the NN2 wire 
along the wire direction is shown in Fig. \ref{bsnn2}. Comparing with Fig.
\ref{Albands} it can be seen that the 3$p$ bands around $E_f$ 
causing the magnetization have lost their degeneracy. However,
in contrast to the unsupported zig-zag wires in Fig. \ref{bs130}
both of the spin-down 3$p$ bands are above the Fermi level.
Therefore the magnetic moments of the wires attain values similar
to those for unsupported linear wires. The energy
bands for NN1 and NN4 geometries show similar features. In the
NN2 geometry the occupied spin-up 3$p$ bands reach the Fermi level
further away from $k_z=0$ than in the NN1 and NN4 geometries. This
explains why the magnetic moment is largest in NN2 geometry.
When the surface-wire distance is reduced, the unoccupied 
spin-down bands touch in the NN1 geometry the Fermi level and
eventually their low-$k_z$ part sink below the Fermi level. This is 
the reason for the gradual decrease of the magnetic moment of the
NN1 wire towards to shorter distances from the surface.

\begin{figure}[hbt]
 \begin{center}
  \epsfig{file=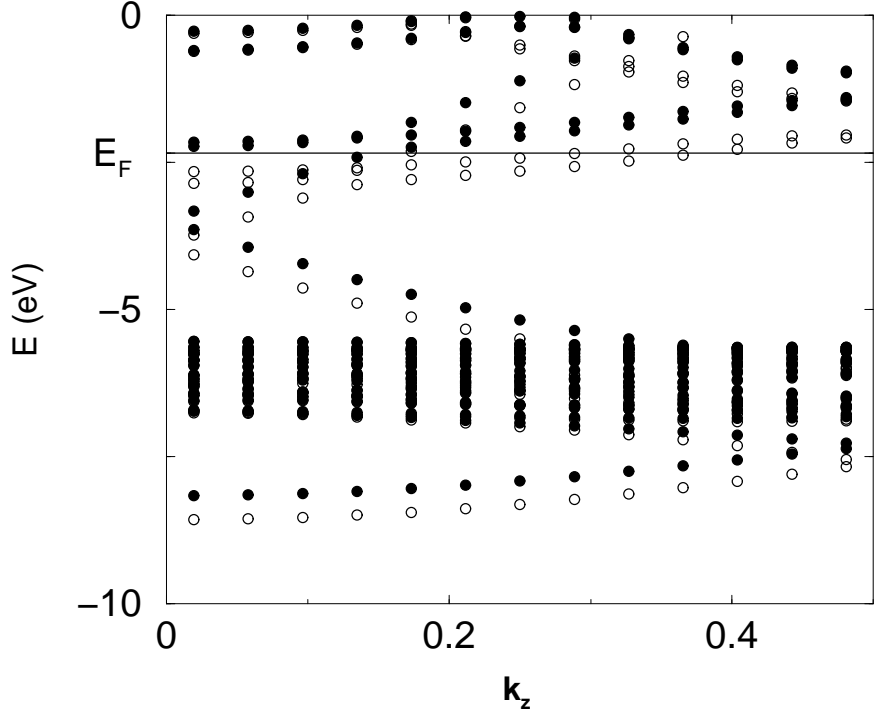,width=0.7\linewidth} 
\caption{{
Band structure of the model Al wire NN2 on NaCl(100) 
(See Fig. {\protect \ref{s-geom} }) along the wire 
direction. Spin-up and spin-down bands are denoted by open and 
filled circles, respectively. The chlorine {\it s}-bands 
included in our calculations are below -10 eV.}}
  \label{bsnn2} \end{center}
\end{figure}

The band structure of the NN2 wire in Fig. \ref{bsnn2} also shows that 
the Na and Cl bands do not disturb the band structure of the Al wire 
near the Fermi level. The Na 3$s$ bands are well above the Fermi 
level. The Cl 3$p$ bands form, at about 6 eV below the Fermi level, 
a  1.5 eV wide region which repells the Al {\it 3s}-bands. 
This repulsion does not affect the Al 3$p$ bands causing the
magnetization. Addition of more NaCl layers into the
supercell causes larger distortions in the Al 3$s$ band, 
but the above discussion about the origin of the magnetization
will not be affected.

In case of the  NN2  wire  we have tested if dimerization could
take place spontaneously. In the test the distance of the wire from 
the surface layer is kept constant corresponding to the previously determined 
equilibrium distance. According to Fig. \ref{w_surf_curves} this
is a fair approximation because the energy minima even for 
the very different geometries occur approximately at the same
distance. Then the two Al atoms in a supercell are simultaneously displaced
symmetrically  in  opposite directions  to  form  a dimerized  wire.
Similarly as in the case of the unsupported linear wires, dimerization  
increases the energy of  the wire. In contrast, a zig-zag deformation
lowers the energy of the NN2 wire as in the unsupported situation.
These notions conform the results obtained for the fully-relaxed Al wire 
on NaCl(100). 

\section{Conclusions}

We have performed density-functional computations to study electronic
and  magnetic properties  of Al nanowires. Unsupported atomic
chains  of  Al become  magnetic  along  elongation  in agreement  with
jellium results.   When relaxing, straight wires  deform to zig-zag
wires  spontaneously.   The  zig-zag  wires exhibit  also  spontaneous
magnetization, although the magnetic moments are lower than those of
straight wires. 

The results for the unsupported wires form the basis
for understanding the properties of nanowires on solid surfaces. 
Many important features of the electronic structure
are conserved when the wire is placed on the solid surface. Metallic
nanowires are soft and the role of the surface is then to provide
a mean to deform in a controlled way the geometry ({\em e.g.}
the bond length) of the nanowire. As a concrete example we have 
considered commensurate Al wires on the NaCl(100) surface. A
relaxation of the Al ion positions results in a zig-zag
geometry and the wire conserves its magnetic moment. The
resulting properties can be understood by studying the band structures 
of model wires on a NaCl layer. 

Our calculations for Al wires on NaCl(100) are among the first studies 
for understanding properties of nanowires on solid surfaces. They
show the importance of relaxing the ionic geometry of the wire. 
In order to gain deeper knowledge more geometries should  be tried. 
For example, an Al wire at the step edge on the  NaCl(100) surface would 
be interesting from the experimental point of view as a way to construct 
a straight wire. Interesting  future topics could also be Al 
wires on surfaces of  different materials, {\it e.g.} on silicon.  

Studies of nanowires on crystal surfaces are of current 
scientific interest  because  such  structures   can  be  implemented  
in  future nanotechnological devices. On the other hand, one-dimensional 
atomic chains on surfaces have been seen during the early stages of the growth
of adsorbate layers and they have been shown to exhibit many interesting
physical phenomena. We hope that our calculations have enlightened
the richness of physical phenomena on these fields and will eventually
encourage experiments searching for magnetism of simple metal nanowires 
on solid surfaces.

\begin{center} 
{\bf ACKNOWLEDGMENTS} \end{center}
 
A. Ayuela is  a Marie Curie Fellow supported by the  EU TMR program 
(Contract No. ERB4001GT954586).
This research has been supported by the Academy of Finland
through its Centre of Excellence Programme (2000 - 2005).
We acknowledge the  generous computing resources  of the Center  for the
Scientific Computing (CSC), Espoo, Finland. 
The authors whish to thank G. Kresse for providing the VASP computer
programs which made this work possible.

%\bibliographystyle{plain}
%\bibliographystyle{prsty}

%\bibliography{omaabbrev,silicide,GMR,dft}
%\bibliography{omaabbrev,paperbib}

%\small

\end{document}